{}
{}

\documentclass[prd,reprint,superscriptaddress,preprintnumbers,nofootinbib]{revtex4-1}

\newif\ifpublic\publictrue

\usepackage{fancyhdr}
\ifpublic\else
\fi
\newif\ifworking\workingtrue

\usepackage{mathtools,mathrsfs,amsbsy,amssymb,latexsym,amsfonts,amscd,
	amsmath,, multirow}
\usepackage{bm}

\usepackage{graphicx}
\usepackage[usenames,dvipsnames]{color}
\usepackage[normalem]{ulem}
\usepackage{natbib}
\usepackage{xcolor}
\usepackage{comment}
\definecolor{linkcolor}{rgb}{0,0,0.6}
\usepackage[ pdftex,colorlinks=true,
pdfstartview=FitV,
linkcolor= linkcolor,
citecolor= linkcolor,
urlcolor= linkcolor,
hyperindex=true,
hyperfigures=false]
{hyperref}

\usepackage{color}
\usepackage{soul,xcolor}
\definecolor{red}{rgb}{1,0,0}
\definecolor{lred}{rgb}{0.3,0,0}
\definecolor{green}{rgb}{0,0.6,0}
\definecolor{blue}{rgb}{0,0,1}
\definecolor{violet}{rgb}{0.8,0,0.8}

\setstcolor{red}
\setul{}{0.8mm}

\def\showkeysrefformat#1{{\normalfont\tiny\ttfamily#1}}
\makeatletter
\def\SK@@ref#1>#2\SK@{%
	{\@inlabelfalse\leavevmode\vbox to\z@{%
			\vss\SK@refcolor\rlap{\vrule\raise .75em%
				\hbox{\showkeysrefformat{#2}}}}}}
\makeatother

\allowdisplaybreaks[3]

\begin{document}
\title{Minimal Massive Supergravity }

\author{Nihat Sadik Deger}
\email{sadik.deger@boun.edu.tr}
\affiliation{Department of Mathematics, Bogazici University, Bebek, 34342, Istanbul, Turkey}
\affiliation{Feza Gursey Center for Physics and Mathematics, Bogazici University, Kandilli, 34684, Istanbul, Turkey}
\affiliation{Erwin Schr\"odinger International Institute for Mathematics and Physics, University of Vienna, Boltzmanngasse 9, 1090, Vienna, Austria}

\author{Marc Geiller}
\email{marc.geiller@ens-lyon.fr}
\affiliation{ENSL, CNRS, Laboratoire de physique, F-69342 Lyon, France}

\author{Jan Rosseel}
\email{jan.rosseel@univie.ac.at}
\affiliation{Faculty of Physics, University of Vienna, Boltzmanngasse 5, 1090, Vienna, Austria}
\affiliation{Division of Theoretical Physics, Rudjer Bo\v{s}kovi\'c Institute, Bijeni\v{c}ka 54, 10000 Zagreb, Croatia}
\affiliation{Erwin Schr\"odinger International Institute for Mathematics and Physics, University of Vienna, Boltzmanngasse 9, 1090, Vienna, Austria}

\author{Henning Samtleben}
\email{henning.samtleben@ens-lyon.fr}
\affiliation{ENSL, CNRS, Laboratoire de physique, F-69342 Lyon, France}
\affiliation{Institut Universitaire de France (IUF)}

\begin{abstract}
Minimal massive gravity in three dimensions propagates a single massive spin-2 mode around an AdS vacuum. It is distinguished by allowing for vacua with positive central charges 
of the asymptotic conformal algebra 
and a bulk graviton of positive energy. We present a new action for the model (and its higher order extensions) in terms of a dreibein and an independent spin connection. From this, we construct its supersymmetric extension. Surprisingly, all vacua complying with bulk and boundary unitarity appear to break supersymmetry spontaneously. In contrast, all supersymmetric vacua have a negative central charge whenever the bulk graviton has positive energy.
\end{abstract}

\maketitle

Three-dimensional (3D) gravity has long been established 
as a distinguished testing ground 
in order to develop, examine, and challenge fundamental issues of 
quantum gravity, black hole physics, and holography,
building on the seminal works \cite{Deser:1983tn,Deser:1983nh,Brown:1986nw,Achucarro:1987vz,Witten:1988hc}.
In what is now understood as a precursor of the Anti-de Sitter/Conformal Field Theory (AdS/CFT) correspondence, 
Brown and Henneaux discovered that Einstein gravity 
on a 3D AdS space
has an infinite-dimensional asymptotic symmetry algebra forming two copies of the Virasoro algebra
with non-vanishing central charges \cite{Brown:1986nw}. This observation has been at the origin of fundamental precision computations \cite{Strominger:1997eq,Birmingham:1998jt}
of the entropy of the Ba\~nados-Teitelboim-Zanelli black hole \cite{Banados:1992wn}, obtained as a discrete quotient of 3D AdS.

Over the years, various 3D gravitational models have been constructed
featuring massive spin-2 excitations around Minkowski and AdS spaces,
including in particular topologically massive gravity (TMG) \cite{Deser:1982vy},
new massive gravity \cite{Bergshoeff:2009hq},
minimal massive gravity (MMG) \cite{Bergshoeff:2014pca},
and  higher order extensions thereof, such as the exotic massive gravity \cite{Ozkan:2018cxj}.

In this letter, we will focus on the MMG model (and its higher order extensions). This model propagates a single massive
spin-2 mode around an AdS background, and is distinguished by the fact that its parameter space allows for a region in which the massive spin-2 mode is neither ghost nor tachyonic, while maintaining both Brown-Henneaux central charges positive.
In this sense, MMG evades the bulk/boundary unitarity clash from which most other 3D massive gravity models suffer, elevating the model to a viable holographic dual of a putative unitary 2D CFT.

To date, none of the 3D bosonic models complying with bulk and boundary unitarity \cite{Bergshoeff:2013xma,Bergshoeff:2014pca}
have been embedded into a supersymmetric theory,
despite the fact that supersymmetry typically provides an invaluable set of tools to establish and corroborate consistent holographic scenarios.
The construction of a supersymmetric extension of MMG is one of the main results of this letter and gives rise to some surprising observations
regarding its vacuum structure, notably an apparent clash between {unbroken} supersymmetry and unitarity for  its AdS vacua.

MMG is defined as a deformation of TMG by terms quadratic in the Riemann tensor according to
\begin{equation}
\frac1{\mu}\,C_{\mu\nu}
+\bar\sigma\,G_{\mu\nu}
+ \bar\Lambda_0\, g_{\mu\nu}
= 
\frac{\gamma}{2\mu^2} \, \epsilon_{\mu\kappa\lambda}  \epsilon_{\nu\sigma\tau} S^{\kappa\sigma} S^{\lambda\tau}  
\;.
\label{MMG0}
\end{equation}
Here, $G_{\mu\nu}$ is the Einstein tensor associated with the metric $g_{\mu\nu}$,
$S_{\mu\nu}$ is the associated Schouten tensor, and $C_{\mu\nu}$ is its Cotton tensor
\begin{align}
G_{\mu\nu}=&\;R_{\mu\nu}-\tfrac12\,g_{\mu\nu}
\;,\qquad
S_{\mu\nu}= R_{\mu\nu}-\tfrac14\,g_{\mu\nu}
\;,\nonumber\\
C_{\mu\nu}=&\;
\epsilon_{\mu\rho\sigma} \nabla^{\rho} S^{\sigma}{}_{\nu}
\;.
\end{align}
In the limit $\gamma\rightarrow0$, equations (\ref{MMG0}) reduce to those of TMG.
The MMG equations (\ref{MMG0}) cannot be derived from a standard action principle of the metric alone. 
In particular, the fact that the r.h.s.\ has zero divergence (as required for consistency) is not automatic, 
but follows on-shell from iterating equation (\ref{MMG0}) itself ---
a mechanism dubbed ``third way consistency'' in \cite{Bergshoeff:2014pca}. 
On the other hand, in \cite{Bergshoeff:2014pca} equations (\ref{MMG0}) could be derived by variation
of a first order Lagrangian with auxiliary fields, in the region of
parameter space where
\footnote{This inequality immediately follows from expressing the parameters of the MMG equations (\ref{MMG0}) in terms of the parameters 
$\{ \sigma,  \alpha, \mu \}$ of the first-order
Lagrangian of \cite{Bergshoeff:2014pca}: $\mu^2\,(1+\gamma\bar\sigma)^2 - {\gamma^3\,\bar\Lambda_0} = \mu^2\,(1+\sigma\,\alpha)^{-2}>0$\,
(recall that $\sigma^2=1$ in \cite{Bergshoeff:2014pca}).}
\begin{equation}
\mu^2\,(1+\gamma\bar\sigma)^2 > \gamma^3\,\bar\Lambda_0
\;.
\label{ktn}
\end{equation}
Absence of a standard action functional has among other things hampered the
construction of the supersymmetric extension of (\ref{MMG0}). As a main result of this letter,
we will employ a new action for MMG (and higher order extensions thereof), in order to construct
the supersymmetric extension of (\ref{MMG0}). As it turns out, the extension is not unique,
and for a given set of coupling constants $\{\gamma, \mu, \bar\sigma, \bar\Lambda_0\}$ we find
(up to four) different supersymmetric extensions of the model. The underlying supersymmetric structures
provide additional tools for the vacuum analysis of the model.

As a starting point, we consider the following class of actions, depending on a dreibein $e_\mu{}^{a}$ and 
an independent (torsionful) spin connection $\varpi_\mu{}^{a}$ as 
\begin{align} \label{LMasterEW}
 {\cal L}[e,\varpi]=&\,
{\cal L}_{0}[e]+ \tau\,\varepsilon^{\mu\nu\rho}\,
    e_\mu{}^{a}{} D[\varpi]_\nu  e_{\rho a} 
 \\
&
+\kappa\,\varepsilon^{\mu\nu\rho}
\left(
\varpi_\mu{}^{a} \partial_\nu \varpi_{\rho a}
+\tfrac13\varepsilon_{a b c}\, \varpi_\mu{}^{a} \varpi_\nu{}^{b} \varpi_\rho{}^{c}
\right)
     \;, \nonumber
\end{align}
with local Lorentz indices $a$, $b$, $\cdots$ (for our conventions see~\footnote{The conventions used in this letter are as follows. We use the $(-++)$ signature convention. The Levi-Civita tensor density and tensor are denoted by $\varepsilon^{\mu\nu\rho}$ and $\epsilon^{\mu\nu\rho}$ respectively. All spinors are Majorana. The covariant derivatives with respect to $\omega$ and $\varpi$ are denoted by $D[\omega]_\mu$ and $D[\varpi]_\mu$. 
On a Lorentz vector $X^a$ and spinor $\Psi$, one has $D[\omega]_\mu X^a = \partial_\mu X^a + \varepsilon^{a b c} \omega_{\mu b} X_c$ and $D[\omega]_\mu \Psi = \partial_\mu \Psi + \frac12 \omega_\mu{}^a \gamma_a \Psi$ and analogously for $D[\varpi]_\mu X^a$ and $D[\varpi]_\mu \Psi$. The torsion and curvature for $\omega$ (and mutatis mutandis for $\varpi$) are defined as $T[\omega]_{\mu\nu}{}^a = 2 D[\omega]_{[\mu} e_{\nu]}{}^a$ and $R[\omega]_{\mu\nu}{}^a = 2\partial_{[\mu} \omega_{\nu]}{}^a + \varepsilon^{a b c} \omega_{\mu b} \omega_{\nu c}$. We have $\gamma_{\mu\nu} = \epsilon_{\mu\nu\rho} \gamma^\rho$ and spinor bilinears obey $\bar{\psi} \chi = \bar{\chi} \psi$ and $\bar{\psi} \gamma_\mu \chi = -\bar{\chi} \gamma_\mu \psi$.}).  
{To begin with, ${\cal L}_{0}[e]$ is an arbitrary gravitational Lagrangian, depending only on the dreibein $e_\mu{}^a$.}
The constants $\tau$ and $\kappa$ denote the coupling constants 
of the torsion and the Chern-Simons term for $\varpi$, respectively.
Variation of the Lagrangian (\ref{LMasterEW}) w.r.t.\ the connection $\varpi$ yields an equation for its curvature
\begin{equation}
R[\varpi]_{\mu\nu, a} 
= - \tfrac{\tau}{\kappa}\,\varepsilon_{a b c}\,
    e_\mu{}^{b}{}   e_{\nu}{}^{c} 
    \;.
    \label{RbEE}
\end{equation}
Variation of (\ref{LMasterEW}) w.r.t.\ the dreibein $e_\mu{}^{a}$ on the other hand
determines the torsion $T[\varpi]_{\mu\nu, a}$ of the connection $\varpi$
by
\begin{equation}
0 = 2\,\mathbb{G}^\mu{}_{a}
+ \tau\,\epsilon^{\mu\nu\rho}\,T[\varpi]_{\nu\rho, a}
\;,
\label{TG}
\end{equation}
where $\mathbb{G}^\mu{}_{a}$ is defined as
\begin{equation}
\delta{\cal L}_0 = \sqrt{-g}\,\mathbb{G}^\mu{}_{a}\,\delta e_\mu{}^{a}
\;.
\label{delL}
\end{equation}
Diffeomorphism and Lorentz symmetry imply that the tensor 
$\mathbb{G}_{\mu\nu}\equiv \mathbb{G}_{\mu}{}^{a}e_{a \nu}$ 
is symmetric and divergence-free
\begin{equation}
\mathbb{G}_{\mu\nu}=\mathbb{G}_{\nu\mu}\;,\qquad
\nabla^\mu\mathbb{G}_{\mu\nu} = 0
\;.
\label{ddG}
\end{equation}
The contorsion of the connection $\varpi$ is defined as
\begin{align}
  K[\varpi]_{\mu}{}^{a} 
=\;&
\varpi_\mu{}^{a}-\omega_\mu{}^{a}
\nonumber\\
=\;&
\tfrac12\,\epsilon^{\rho\sigma\tau} \left(
e_\mu{}^{b}e_\rho{}^{a}  -\tfrac12 e_\mu{}^{a}  e_\rho{}^{b}\right)
T[\varpi]_{\sigma\tau, b}
\;,
\end{align}
in terms of the torsionless Levi-Civita connection $\omega_\mu{}^{a}$. The field equations (\ref{TG})
can then be rewritten as
\begin{equation}
K[\varpi]_{\mu}{}^{a} = -\frac1{\tau}\left(\mathbb{G}_{\mu}{}^{a} - \tfrac12e_{\mu}{}^{a}\,\mathbb{G} \right)
\equiv -\frac1{\tau}\, \mathbb{S}_{\mu}{}^{a}
\;,
\label{KS}
\end{equation}
where $\mathbb{G}\equiv \mathbb{G}_\mu{}^{a}e_{a}{}^\mu$\,. Finally, using the general relation
between curvature and contorsion
\begin{align}
R[\varpi]_{\mu\nu}{}^{a}
=&\, R[\omega]_{\mu\nu}{}^{a} +2\,D[\omega]_{[\mu}   K[\varpi]_{\nu]}{}^{a}
\nonumber\\&
+\varepsilon^{a b c}\,K[\varpi]_{\mu b} K[\varpi]_{\nu c}
\;,
\label{RRKK}
\end{align}
and combining this with (\ref{RbEE}) and (\ref{KS}), yields the equation
\begin{equation}
\epsilon_{\mu\sigma\tau} \,\nabla^{\sigma}   \mathbb{S}^{\tau}{}_\nu  
-\tau\,G_{\mu\nu}
+\tfrac{\tau^2}{\kappa}\,g_{\mu\nu}
=
\tfrac1{2\tau}\,\epsilon_{\mu\sigma\tau}  \epsilon_{\nu\kappa\lambda}\,\mathbb{S}^{\sigma\kappa}\mathbb{S}^{\tau\lambda}\,,
\label{eom_final}
\end{equation}
exclusively formulated in terms of the dreibein, 
with $\mathbb{S}_{\mu\nu}$ defined via (\ref{KS}), (\ref{delL}).
Just as in (\ref{MMG0}), these equations are ``third-way consistent'' in that the vanishing of the
divergence of the r.h.s.\  follows from iterating equation (\ref{eom_final}) itself
(together with the relations (\ref{ddG})).
A key feature of this construction is that the final equations (\ref{eom_final}) do not arise directly among the 
Euler-Lagrange equations, but only after combination with the integrability conditions (\ref{RRKK}).\footnote{ 
A similar mechanism has been employed in \cite{Deger:2022znj} for a description
of the so-called third way consistent deformation of Yang-Mills theory in terms of a gauged scalar sigma model.}

Let us also point out that consistent matter couplings to (\ref{eom_final})
are straightforwardly implemented in the Lagrangian (\ref{LMasterEW}) by 
\begin{align}
{\cal L}_0[e] & \longrightarrow\; {\cal L}_0[e] + {\cal L}_{\rm matter}[e, \dots]
\;,
\nonumber\\
\Longrightarrow\quad
\mathbb{G}_{\mu\nu} 
& \longrightarrow \;\mathbb{G}_{\mu\nu} + T_{\mu\nu}
\;,
\end{align}
with the standard (covariantly conserved) energy-momentum tensor $T_{\mu\nu}$,
reproducing the (on-shell) results of \cite{Arvanitakis:2014yja,Ozkan:2018cxj}.

In the following, we specialize to MMG, by setting
\begin{equation}
{\cal L}_0[e] =
\frac1{G_3}\,\varepsilon^{\mu\nu\rho}e_\mu{}^{a}R_{\nu\rho, a}
+\lambda\varepsilon^{\mu\nu\rho}\varepsilon_{a b c} e_\mu{}^{a} e_\nu{}^{b} e_\rho{}^{c}
\;,
\label{L0MMG}
\end{equation}
where for simplicity we will set the gravitational constant $G_3=1$. The resulting equations (\ref{eom_final})
then reproduce (\ref{MMG0}), with the (bijective up to rescaling) translation of parameters according to
\begin{equation}
\frac{\mu}{\gamma} = {\tau}\;,\;\;\;\;
\mu\bar\Lambda_0 = \frac{9\lambda^2}{4\tau}+\frac{\tau^2}{\kappa} \;,\;\;\;\;
\mu\bar\sigma =-\tau -\frac{3\lambda}{2\tau}
\;.
\label{param}
\end{equation}
Equivalently, (\ref{L0MMG}) can be replaced by the first order Palatini Lagrangian in terms of an independent connection $\omega$,
\begin{equation}
{\cal L}_0[e,\omega] =
\varepsilon^{\mu\nu\rho}\Big(e_\mu{}^{a}R[\omega]_{\nu\rho, a}
+\lambda \varepsilon_{a b c} e_\mu{}^{a} e_\nu{}^{b} e_\rho{}^{c}\Big)\,,
\label{L0MMG_Pal}
\end{equation}
such that the final Lagrangian (\ref{LMasterEW}) is given by the sum of the so-called ``standard'' and the ``exotic''
action of 2+1 gravity \cite{Witten:1988hc}, however with both actions carrying different spin connections $\omega$, and $\varpi$, respectively
\footnote{The appearance of two independent spin connections is responsible for the presence of a massive degree of freedom 
in this formulation of MMG.
For $\omega=\varpi$, the Lagrangian (\ref{LMasterEW}) would merely correspond to a reformulation of 3D
(topological) gravity, such as studied in \cite{Mielke:1991nn,Cacciatori:2005wz}}.
In this first order formulation, and after redefinition
\begin{equation}
\omega_\mu{}^{a} = \Omega_\mu{}^{a}+ \alpha\,h_\mu{}^{a}    
\;,
\;\;
\varpi_\mu{}^{a} = \Omega_\mu{}^{a} +\frac{\sqrt{-\kappa\tau}}{\kappa}\,e_\mu{}^{a}
\;,
\end{equation}
the Lagrangian (\ref{LMasterEW}), 
{ written as ${\cal L}[e,\Omega,h]$,}
reproduces the first-order Lagrangian of \cite{Bergshoeff:2014pca}.
The condition $\kappa\tau<0$ precisely defines the region in parameter space (\ref{ktn}) in which
the Lagrangian of \cite{Bergshoeff:2014pca} exists.

We will now use the Lagrangian (\ref{LMasterEW}), (\ref{L0MMG}) in its second order form
as a starting point for the construction of supersymmetric extensions of the model.
Separate supersymmetrization of the two parts of (\ref{LMasterEW}) is known
in terms of super Chern-Simons theories,
with in particular (\ref{L0MMG}) admitting a general ${\cal N}=(p,q)$ supersymmetric extension~\cite{Achucarro:1987vz}.
Again, the non-trivial structure here arises since both parts of (\ref{LMasterEW}) share the same dreibein $e_\mu{}^{a}$
while carrying independent spin connections $\omega$, and $\varpi$.
For simplicity, we will only attempt to impose minimal (${\cal N}=(1,0)$) supersymmetry.

Our ansatz for the fermionic sector of the model carries two gravitino fields, $\psi_\mu$ and $\chi_\mu$, sharing only 
one local supersymmetry, with spinor parameter $\epsilon$. 
Reminiscent of the first order formulation of TMG \cite{Sezgin:2009dj,Routh:2013uc},
this allows the first order fermionic Lagrangian 
(expected for the supersymmetrization of (\ref{LMasterEW}), (\ref{L0MMG}))
to consistently accommodate a massive spin-3/2 mode as a superpartner to the massive spin-2 mode.
Indeed, this turns out to be the correct structure. 
{Leaving the technical details for \cite{Deger:2022xxx}, let us only spell out the result.}
To quadratic order in the fermions, their Lagrangian is given by
\begin{align}
{\cal L}_{\rm ferm} =\;&
\frac1{\zeta^2}\,\varepsilon^{\mu\nu\rho}
 \big((1+\zeta)\bar\psi{}_\mu+\bar\chi_\mu \big) D[\omega]_\nu \big((1-\zeta)\psi{}_\rho+\chi_\rho\big)
\nonumber\\
&{}
 +\frac1{2\zeta^2}\, \varepsilon^{\mu\nu\rho}\,K[\varpi]_\nu{}^{a} 
 \big(\bar\psi_\mu+\bar\chi_\mu \big) \gamma_{a} \big(\psi_\rho+\chi_\rho \big)
\nonumber\\
&{}
  -\tau\,\varepsilon^{\mu\nu\rho}
 \,\bar\psi{}_\mu  \gamma_{\nu} \chi_\rho 
  +\frac12\,\tau\,\varepsilon^{\mu\nu\rho}
 \,\bar\chi{}_\mu \gamma_{\nu} \chi_\rho 
\nonumber\\
&{}
  +\frac{1}4\left((\zeta^2-2)\,\tau+\frac1{\zeta^2\,\kappa}\right)
  \varepsilon^{\mu\nu\rho}
 \,\bar\psi{}_\mu \gamma_{\nu} \psi_\rho 
\;,
\label{Lf_exp}
\end{align}
with the couplings determined as functions of a new parameter $\zeta$\,.
{The full} Lagrangian can be shown to be invariant under the
supersymmetry transformations (also given to lowest order in the fermions)
\begin{align}
\delta e_\mu{}^{a} =\;&
\frac{1}{2}\,\bar\psi_\mu \gamma^{a} \epsilon
\;,\nonumber\\
\delta \varpi_\mu{}^{a} =\;&
-\frac1{2\zeta^2\kappa}\left(\bar\psi_\mu+\bar\chi_\mu \right)  \gamma^{a} \epsilon
-\frac{1}{2}\,D[\varpi]_\mu \left(\bar\chi_\nu \epsilon \,e^{\nu a}\right)
\;,\nonumber\\
\delta \psi_\mu =\;& D[\omega]_\mu \epsilon 
-\frac14 \left(\zeta^2\tau+\frac1{\zeta^2\,\kappa}\right) \gamma_\mu \epsilon
\;,\nonumber\\
\delta \chi_\mu =\;&  \frac12\,K[\varpi]_\mu{}^{a} \gamma_{a} \epsilon
-\frac14 \left(\zeta^2\tau-\frac1{\zeta^2\,\kappa}\right) \gamma_\mu \epsilon
\;,
\label{susy_exp_mod}
\end{align}
up to quartic terms in the fermions, which can be removed by higher order fermion contributions
to the Lagrangian and transformation rules.\footnote{Thus pushing potential obstacles to sixth order in the fermions.} 
Details will appear elsewhere \cite{Deger:2022xxx}.

{The final result is thus given by the sum of the bosonic Lagrangian (\ref{LMasterEW}), (\ref{L0MMG}) and the
fermionic Lagrangian (\ref{Lf_exp}).}
Supersymmetry requires the following relation 
\begin{equation}
\lambda=
\frac{1+\zeta^2 \,\kappa \tau  \left(2 \,\zeta ^2+\left(\zeta^2-4\right) \zeta ^4 \,\kappa \tau
   +4\right)}{12 \,\zeta^4 \, \kappa^2}
\;,
\label{lambda_alpha}
\end{equation}
between the parameter $\zeta$ parametrizing the fermionic couplings,
and the coupling constants $\{\lambda, \kappa, \tau\}$ of the bosonic model (\ref{LMasterEW}), (\ref{L0MMG}).
Supersymmetrizability of the MMG model thus translates into the existence of real roots (for $\zeta$) of (\ref{lambda_alpha}).
A necessary and sufficient set of conditions for the existence of such real roots is
\begin{align}
\bullet\;\;\;
&
 \frac{3\lambda}{\tau^2}-\frac{1}{\kappa\tau}+1 \ge 0 \;,
\nonumber\\
\bullet\;\;\;
&
\mbox{either}\;\;\,\kappa\tau\ge-1\;\;\mbox{or}\;\;\,
\frac{3\lambda}{\tau^2} \ge -\frac{2}{\sqrt{-\kappa\tau}}
\;.
\label{cond_vac}
\end{align}
There are in general up to eight real roots (pairwise related by the flip $\zeta\rightarrow-\zeta$).
{While this may appear to place strong constraints on the model,
remarkably our analysis below reveals that every MMG model (\ref{MMG0}) admitting an AdS vacuum
also admits a supersymmetric extension.}

The bosonic MMG equations (\ref{MMG0}) are obtained from the second order Lagrangian 
(\ref{LMasterEW}), (\ref{L0MMG}), after elimination of the connection $\varpi$ by its field equations. We can 
now carry out the analogue construction in the fermionic sector. The fermionic field equations,
obtained from variation of (\ref{Lf_exp})
\begin{align}
 D[\omega]_{[\mu}\psi_{\nu]}
 = \frac14\left(\zeta^2\,\tau+\frac{1}{\zeta^2\,\kappa}\right) \gamma_{[\mu} \psi_{\nu]}
 - \tau\,\gamma_{[\mu} \chi_{\nu]}
 \;,
\end{align}
can be solved algebraically for $\chi_\mu$.
Plugging this solution back into the remaining fermionic field equations, we find after some computation
(still to lowest order in fermions)
\begin{align}
\tau\,C^\rho=&\,
\frac1{8}\,\nu(2)\nu(-2)
\,R^{\rho}
-\frac1{4}\,\nu(0)\, \epsilon^{\mu\nu\rho}
\,\gamma_\sigma \psi_\nu\, S_\mu{}^\sigma \,
\nonumber\\
&{}
-\frac{1}{32}
\nu(0)\nu(2)\nu(-2)\,
 \epsilon^{\mu\nu\rho}\, \gamma_{\mu} \psi_\nu
-\frac1{2}
\, R^{\mu}\, G_\mu{}^\rho \,
\nonumber\\
&{}
-\frac1{2}\,\epsilon^{\rho\mu\nu}
\,\gamma^\sigma R_\mu\, G_{\nu\sigma} \,
-\frac1{2}\,\epsilon^{\rho\mu\nu}
\gamma_\mu   R^{\sigma}\, G_{\nu\sigma} \,
\;,
\label{eom_fermion_psi}
\end{align}
where we have 
 introduced the gravitino curvature
\begin{equation}
R^\mu = \epsilon^{\mu\nu\rho}\,D[\omega]_\nu \psi_\rho
\;,
\end{equation}
and the ``Cottino vector-spinor'' \cite{Gibbons:2008vi}
\begin{equation}
C^\mu = 
\gamma^\rho\gamma^{\mu\nu}\,D[\omega]_\nu R_\rho - \epsilon^{\mu\nu\rho}\,S_{\rho\sigma}\,\gamma^\sigma \psi_\nu
\;,
\end{equation}
and moreover defined the functions
\begin{equation}
\nu(n) \equiv (\zeta+n)^2\,\tau+\frac{1}{\zeta^2\,\kappa}
\;.
\end{equation}
We thus obtain the second-order fermionic field equation, entirely expressed in terms of the dreibein $e_\mu{}^{a}$ 
and the gravitino $\psi_\mu$.
Equation (\ref{eom_fermion_psi}) constitutes the ``super-partner'' to the bosonic MMG equations (\ref{MMG0}).
{In the process, the latter receive additional source terms bilinear in the fermions. In order not to spoil consistency of the field equations  (\ref{MMG0}), these source terms have to satisfy certain consistency conditions which in turn are implied by (\ref{eom_fermion_psi}). Details will appear in \cite{Deger:2022xxx}.}

Let us also note that in the TMG limit, where the r.h.s.\ of (\ref{MMG0}) vanishes, equation (\ref{eom_fermion_psi}) 
reduces to
\begin{equation}
C^\rho=
-2\,\mu \bar\sigma R^{\rho} 
-\frac{\mu \bar\sigma}{\sqrt{-\bar\Lambda_0}}  \,\gamma^{\rho\nu}
\psi_\nu
\;.
\label{super-TMG}
\end{equation}
This is precisely
the super-TMG equation from~\cite{Gibbons:2008vi} (where $\bar\sigma=1$ was assumed).

Let us now explore the landscape of (A)dS vacua of the super-MMG model
{and in particular localize the bulk/boundary unitary AdS vacua discovered in \cite{Bergshoeff:2014pca}}. 
The bosonic MMG equations (\ref{MMG0})
admit maximally symmetric vacua
$G_{\mu\nu} =  -\Lambda\,g_{\mu\nu}$,
provided that
\begin{equation}
\mu^2\,\bar\sigma^2 \ge \gamma\,\bar\Lambda_0
\;.
\end{equation}
With the translation of parameters (\ref{param}), this turns out to precisely coincide with the first condition in (\ref{cond_vac}).
The cosmological constants are given by
\begin{equation}
\Lambda_\pm = -\tau^2 \Big(
 (1\pm \Gamma)^2 + \frac{1}{\kappa\tau} \Big)
 ,\;\;\;
 \Gamma \equiv \sqrt{\frac{3\lambda}{\tau^2}-\frac{1}{\kappa\tau}+1}
 \,.
 \label{Lpm}
\end{equation}
As a consequence, every supersymmetrizable MMG model admits maximally symmetric vacua.
Evaluating the super-MMG field equations (\ref{MMG0}) with (\ref{param}) and (\ref{lambda_alpha}), we 
obtain the values of their cosmological constants as
\begin{equation}
\begin{array}{l}
\displaystyle{\Lambda_{\rm susy} =
 -\frac{(1+\zeta^4\,\kappa\tau)^2}{4\,\zeta^4\,\kappa^2}}
  \equiv -\frac1{\ell^2}
\;,\\[2ex]
\displaystyle{\Lambda_{\rm ns} =
-\frac{1+\zeta^2\,\kappa\tau\left(8+2\,\zeta^2+\zeta^2\,(\zeta^2-4)^2\,\kappa\tau\right)}{4\,\zeta^4\,\kappa^2}}
\;.
\end{array}
\label{sns}
\end{equation}
The first vacuum in (\ref{sns}) is AdS (or Minkowski) and preserves {part of the} supersymmetry with the 
Killing spinor defined by
\begin{equation}
D[\omega]_\mu \epsilon 
-\frac1{2\ell}\,\gamma_\mu \epsilon =0 \, .
\end{equation}
From (\ref{susy_exp_mod}), it then follows that $\delta\psi_\mu=0$ is satisfied as usual for AdS,
whereas $\delta \chi_\mu=0$ holds identically, as a consequence of \eqref{KS}. On the other hand, for 
the non-supersymmetric vacuum $\Lambda_{\rm ns}$  in (\ref{sns}), the Killing spinor equations for 
$\psi_\mu$ and $\chi_\mu$  \eqref{susy_exp_mod} cannot be solved simultaneously.

Linearizing and diagonalizing the bosonic Lagrangian around the supersymmetric AdS vacuum $\Lambda_{\rm susy}$
(with $\tilde{e}_\mu{}^a$ as background dreibein) yields 
\begin{align}
{\cal L}_{\rm lin}=\;&
\varepsilon^{\mu\nu\rho}
\,
\alpha_+\,
f^{(+)}_\mu{}^{a}
\left(
\partial_\nu f^{(+)}_{\rho,a}
-\ell^{-1}\, \varepsilon_{a b c}\,  f^{(+)}_\nu{}^{b}\, \tilde{e}_\rho{}^{c}
\right)
\nonumber\\
&{}
+\varepsilon^{\mu\nu\rho}
\,
\alpha_-\,
f^{(-)}_\mu{}^{a}
\left(
\partial_\nu f^{(-)}_{\rho,a}
+\ell^{-1}\, \varepsilon_{a b c}\,  f^{(-)}_\nu{}^{b}\, \tilde{e}_\rho{}^{c}
\right)
\nonumber\\
&{}
-\varepsilon^{\mu\nu\rho}
\,
\alpha_0\,
p_\mu{}^{a}
\left(
\partial_\nu p_{\rho,a}
+M\, \varepsilon_{a b c}\,  p_\nu{}^{b}\, \tilde{e}_\rho{}^{c}
\right)
\;,
\end{align}
exhibiting two massless and one massive spin-2 mode, $f^{(\pm)}$ and $p$, respectively,
with coefficients factorizing in terms of $\zeta$ in an intriguing pattern as
\begin{align}
\alpha_+ =\;&
\kappa\,(1-\zeta^2)\,(1+\zeta^4\,\kappa\tau)
\;,
\nonumber\\
\alpha_-=\;&
\kappa\,(1+\zeta^2\,\kappa\tau)(1+\zeta^4\,\kappa\tau)
\;,
\nonumber\\
\alpha_0=\;&
\kappa\,(1-\zeta^2)\,(1+\zeta^2\,\kappa\tau)
\;,
\label{alpha3}
\end{align}
and with the mass given by
\begin{equation}
M\ell = 
-\frac{(1+\zeta^2\,\kappa\tau)+\zeta^2(1-\zeta^2)\kappa\tau}{1+\zeta^4\,\kappa\tau}
\;.
\end{equation}
The no-tachyon condition $M^2\ell^2>1$ for the massive \hbox{spin-2} mode translates into
\begin{equation}
\zeta^2 \kappa\tau\,(1-\zeta^2)\,(1+\zeta^2\,\kappa\tau) > 0
\;.
\end{equation}
An analysis following \cite{Bergshoeff:2014pca,Bergshoeff:2019rdb} quickly shows that
imposing no-ghost and no-tachyon unitarity conditions on the massive spin-2 mode necessarily
implies negative central charges. Remarkably, all supersymmetric AdS vacua thus exhibit the clash between
bulk and boundary unitarity.

\begin{figure}[h]
   \centering
   \includegraphics[width=9cm]{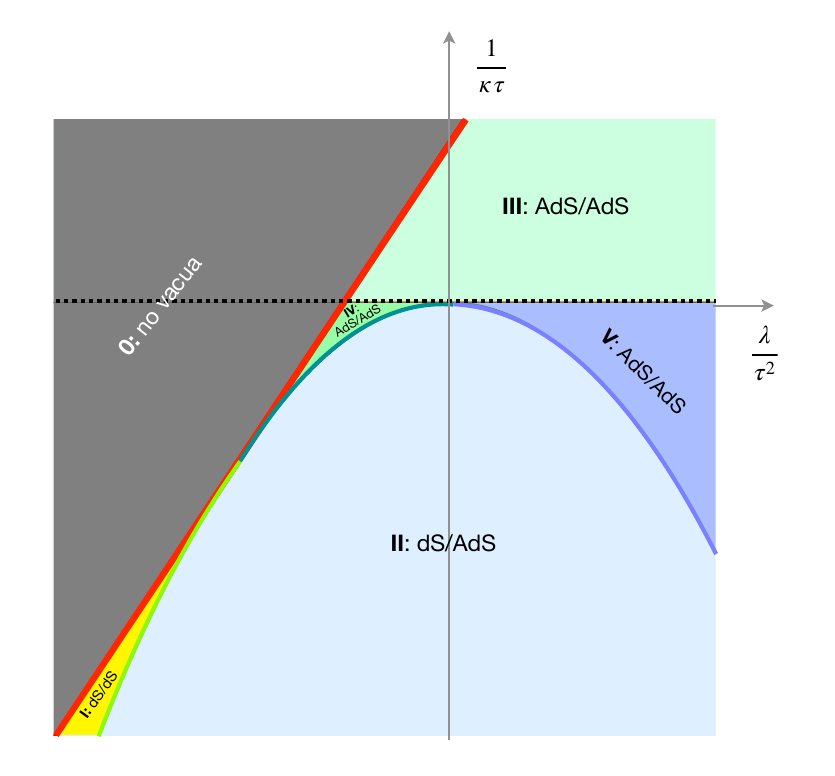}
   \caption{Different regions in parameter space. As long as the MMG model admits an AdS vacuum,
   i.e. outside of the gray and yellow areas, it admits at least one supersymmetric extension.
   The AdS vacua evading the bulk/boundary unitarity clash~\cite{Bergshoeff:2014pca} 
   are the non-supersymmetric AdS vacua in region V.}
   \label{fig:dSAdS}
\end{figure}

We can finally map out the full landscape of (A)dS vacua in order to reconcile these results with the earlier 
findings of \cite{Bergshoeff:2014pca}. To this end, we combine the conditions on supersymmetrizability (\ref{cond_vac})
with the values of the cosmological constants (\ref{Lpm}), (\ref{sns}) in order to identify the various regions in parameter space
as depicted in figure~\ref{fig:dSAdS}.

\begin{enumerate}

\item[{\bf 0}:] In this region
\begin{equation}
\frac1{\kappa\tau} > 1+\frac{3\lambda}{\tau^2}
\;,
\end{equation}
the first condition in (\ref{cond_vac}) is violated. According to the above discussion,
no (A)dS vacua exist in this region and the model is not supersymmetrizable.

\item[{\bf I}:] 
This region is bounded by region 0 and the parabola 
\begin{equation}
\frac1{\kappa\tau} = -\frac94\,\Big(\frac{\lambda}{\tau^2}\Big)^2
\;,
\label{parab}
\end{equation}
with $\frac{\lambda}{\tau^2}<-\frac23$\,.
The second condition in (\ref{cond_vac}) is violated, thus the model is not supersymmetrizable.
With (\ref{Lpm}), it follows that both vacua $\Lambda_\pm$ are of dS type.

\item[{\bf II}:] This region is bounded from above by the parabola (\ref{parab}). $\Lambda_+$ is an AdS vacuum, $\Lambda_-$ is dS. 
There are two different solutions to (\ref{lambda_alpha}), i.e.\ two supersymmetric extensions of the bosonic model, 
in both of which $\Lambda_+$ is supersymmetric, but exhibits the bulk/boundary unitarity clash.

\item[{\bf III}:] This region is not covered by the Lagrangian of~\cite{Bergshoeff:2014pca},
since $\kappa\tau>0$.
Both vacua $\Lambda_\pm$ are of AdS type. Equation (\ref{lambda_alpha}) admits two
different solutions
\begin{equation}
{\zeta_\pm^2} = 
(1\pm \Gamma) +  \sqrt{(1\pm \Gamma)^2+\frac1{\kappa\tau}}\;.
\end{equation}
There are thus two supersymmetric extensions  of the bosonic model, satisfying
\begin{equation}
\Lambda_{\rm susy}(\zeta_\pm) = \Lambda_\pm = \Lambda_{\rm ns}(\zeta_\mp)
\;.
\label{Lsns}
\end{equation}
That is, each of the vacua (\ref{Lpm}) of the bosonic model is supersymmetric in one of the supersymmetric extensions 
and non-supersymmetric in the other.
Both vacua exhibit the bulk/boundary unitarity clash.

\item[{\bf IV}:] Both vacua $\Lambda_\pm$ are AdS.
There are four supersymmetric extensions of the model with a structure similar to (\ref{Lsns}), i.e.\
each of the AdS vacua of the bosonic model is supersymmetric in some supersymmetric extension(s) of the model.
Again, both vacua exhibit the bulk/boundary unitarity clash.

\item[{\bf V}:] Both vacua $\Lambda_\pm$ are AdS.
There are two solutions of equations (\ref{lambda_alpha})
\begin{equation}
{\zeta_\pm^2} = 
(1+ \Gamma) \pm  \sqrt{(1+ \Gamma)^2+\frac1{\kappa\tau}}\;,
\end{equation}
i.e.\ two supersymmetric extensions of the model. In both of these, the vacuum $\Lambda_+$ is supersymmetric, 
whereas $\Lambda_-$ is not. A careful translation of parameters shows that this region is hosting all the vacua identified 
in \cite{Bergshoeff:2014pca} as evading the bulk/boundary clash. 
Specifically, these are the non-supersymmetric $\Lambda_-$ vacua,
as is also consistent with the above analysis of supersymmetric vacua.

\end{enumerate}

Along the border lines separating the different regions in figure~\ref{fig:dSAdS}, there is always one Minkowski vacuum together with an (A)dS vacuum. The red line is the so-called merger line~\cite{Arvanitakis:2014yja} in which both (A)dS vacua of the model coincide. Note that the two regions $\kappa\tau < 0$ and $\kappa\tau > 0$ of the parameter space are not connected, as the model (\ref{LMasterEW}) degenerates for $\kappa\tau=0$ (or $\frac1{\kappa\tau}=0$).

As anticipated above, the analysis shows that MMG admits supersymmetric extensions in all regions (II.--V.) that
admit AdS vacua. In particular, the bulk/boundary unitary AdS vacua discovered in \cite{Bergshoeff:2014pca} are all situated
in region V. and can be embedded into supersymmetric models. Moreover, the analysis shows that in all these vacua 
supersymmetry is spontaneously broken.

The supersymmetric extensions of the MMG model thus offer new perspectives on the AdS vacuum analysis of the bosonic model
and it will be most interesting to explore its repercussions in the context of other solutions of the model, such as~\cite{Arvanitakis:2014yja}.
The product pattern of the coefficients (\ref{alpha3}) in terms of the supersymmetry parameter $\zeta$
indicates the location of the chiral points at which one of the central charges vanishes.
The underlying supersymmetric structure will also be of great value for the unitarity analysis at these special points.
Further interesting research directions include the construction of possible supersymmetric matter couplings to MMG, as well as the supersymmetrization of
higher order extensions of the model \cite{Ozkan:2018cxj,Afshar:2019npk} --- given that our action (\ref{LMasterEW}) naturally accommodates all such generalizations.
A superspace formulation of our construction would be highly desirable to address these issues.

Interestingly, we have identified different minimal supersymmetric  extensions of the same bosonic MMG in which different vacua of the bosonic model appear supersymmetric. This may be read as a hint of an underlying structure of extended supersymmetry into which these models could be embedded, as is typical for such twin supergravities \cite{Roest:2009sn}. As a technical challenge this would require to embed the single massive bosonic degree of freedom into some extended multiplet structure.

Perhaps the most surprising result of our analysis is the observed clash between supersymmetry and bulk/boundary unitarity.
It is precisely the AdS vacua in which supersymmetry is spontaneously broken which reconcile positive central
charges with a positive energy bulk graviton.
It would be very interesting to understand if this observed clash of {unbroken} supersymmetry and unitarity goes back to some 
more fundamental principle and has deeper implications for holography.
The minimal massive supergravity constructed in this paper provides a natural starting point
for studying aspects of holography around non-supersymmetric vacua.
The simultaneous (and unavoidable) presence of a supersymmetric AdS vacuum with bulk/boundary unitarity clash and
a second AdS vacuum avoiding the clash but breaking supersymmetry allows to probe such issues in a single model.

A more detailed version of the presented results will appear elsewhere \cite{Deger:2022xxx}.

\section*{Acknowledgements}
We would like to thank Olaf Hohm {and Mehmet Ozkan} for useful discussions and comments. JR and NSD are 
grateful to the Erwin Schr\"odinger Institute (ESI), Vienna where part of this work was done in the framework of the ``Research in Teams'' Programme. NSD also wishes to thank ENS de Lyon for hospitality during the course of this work.


%

\end{document}